\documentclass[12pt]{elsart}
\usepackage{amsmath,amssymb}
\usepackage{graphicx,psfrag,here,epsfig}
\usepackage{pstricks}
\usepackage{pst-node}
\usepackage{epsf}
\usepackage{citesort}
\allowdisplaybreaks[1]

\begin{document}
%%%%%%%%%%%%%%%%%%%%%%%%%%%%%%%%%%%%%%%%%%%%%%%%%%%%%%%%%%%%%%%%%%%%%%%%%%%%%
\newcommand{\co}{\; \; ,}
\def\words#1{\mbox{\small{\,#1}}}
\def\bea{\vskip-4mm\begin{eqnarray}}
\def\eea{\end{eqnarray}\vskip-3mm}
\def\eq{\vskip-4mm\begin{eqnarray}}
\def\en{\end{eqnarray}\vskip-3mm}
\def\be{\begin{equation}}
\def\ee{\end{equation}}
\newcommand{\ed}{\end{document}}

\newcommand{\nnnl}{\nonumber\\}
\newcommand{\nn}{\nnnl}
\newcommand{\fs}{\, . \,}

\newcommand{\mn}{M_{\pi^0}}
\newcommand{\mc}{M_{\pi}}
\newcommand{\dpi}{\Delta_{\pi}}
\newcommand{\kl}{K_L}

\newcommand{\kd}{K^\dagger}
\newcommand{\phin}{\Phi_0}
\newcommand{\phip}{\Phi_+}
\newcommand{\phim}{\Phi_-}
\newcommand{\pin}{\pi^0}

\def\query#1{\marginpar{\begin{flushleft}\footnotesize#1\end{flushleft}}}%

\runauthor{Bissegger, Fuhrer, Gasser, Kubis, Rusetsky}
\runauthor{Kubis, Rusetsky}
\renewcommand{\theequation}{\arabic{equation}}
\begin{frontmatter}

\begin{flushright}
HISKP--TH--07/24
\end{flushright}

\title{\Large\bf Cusps in \boldmath{$\kl\rightarrow 3\pi$} decays}

\author[Bern]{M.~Bissegger},
\author[Bern]{A.~Fuhrer},
\author[Bern]{J.~Gasser},\hspace{4cm}
\author[Bonn]{B.~Kubis}, 
\author[Bonn]{A.~Rusetsky\thanksref{Tbilisi}}

\address[Bern]{Institute for Theoretical Physics, University of Bern,
Sidlerstr. 5, CH--3012 Bern, Switzerland}
\address[Bonn]{Helmholtz--Institut f\"ur Strahlen-- und Kernphysik,
Universit\"at Bonn, Nussallee~14--16, D--53115 Bonn, Germany}

\thanks[Tbilisi]{On leave of absence from:
High Energy Physics Institute, Tbilisi State University,
University St.~9, 380086 Tbilisi, Georgia.}

\begin{abstract}
  The pion mass difference generates a pronounced cusp in
  $K\rightarrow3\pi$ decays, the strength of which is related to
  the $\pi\pi$ $S$--wave scattering lengths.
  We apply 
  an effective field theory framework developed earlier to evaluate
  the amplitudes for $\kl\to 3\pi$ decays in a systematic manner,
  where the strictures imposed by analyticity and unitarity are respected
  automatically. The amplitudes for the decay $\eta\to 3\pi$ are also given.
\end{abstract}


\end{frontmatter}

\noindent{\bf 1.}
The investigation of the so--called cusp effect in $K^+ \to \pi^+\pi^0\pi^0$
 decays has become a fully competitive method 
for the extraction of the $S$--wave $\pi\pi$ scattering lengths
from experimental data.
Following refined versions of the original proposal by 
Cabibbo~\cite{Cabibbo:2004gq,Cabibbo:2005ez,CGKR},
the combination $a_0-a_2$ has been determined from very high statistics 
data~\cite{Batley,dilellakaon07} to an accuracy
mainly limited by remaining shortcomings in the theoretical description of the decay amplitudes.
Missing ingredients are in particular (real and virtual) photon corrections. Here, an important 
step has recently been performed by Isidori~\cite{isidorirad}, who has evaluated radiative 
corrections in multi--body meson decays, in particular, 
for the fully charged channel $K^+\to \pi^+\pi^+\pi^-$, in the soft photon approximation.
Once these corrections  are available in all channels,  $K\to 3\pi$ decays,
combined with the information gained from $K_{e4}$ 
decays~\cite{Pislak:2003sv,blochkaon07}
 and the pionium lifetime~\cite{Adeva:2005pg}, have the potential to 
test the very precise theoretical prediction
of the scattering lengths~\cite{Colangelo:2000jc,Colangelo:2001df} experimentally.
 For recent phenomenological 
determinations of the scattering lengths, we refer the reader to 
Refs.~\cite{descotes,yndurain,kaminski}.

As the strong impact of the unitarity cusp near the $\pi^+\pi^-$ threshold
is a universal feature of the $\pi^0\pi^0$ scattering amplitude~\cite{MMS},
it is present also in other decays, like $K_L \to 3\pi^0$, $\eta\to 3\pi^0$ etc.
The strength of the cusp in $K_L\to 3\pi^0$ 
is  reduced by about an order of magnitude compared to $K^+ \to \pi^+\pi^0\pi^0$,
hence the experimental situation in order to gain information on $\pi\pi$ scattering
lengths is far less favourable~\cite{dilellakaon07}.
However, the motivation to study this channel all the same is twofold:  
firstly, experimental efforts to at least see the cusp are under way~\cite{dilellakaon07}; secondly, 
the $K_L \to 3\pi^0$ system provides an excellent object for exploratory
studies of the most important electromagnetic effects in the cusp region, before immersing oneself
into the even more relevant, but simultaneously more difficult case of $K^+$ decays.

The $K_L\to 3\pi$ decays have been studied with regard to the cusp phenomenon
before.  Ref.~\cite{Cabibbo:2005ez} uses unitarity, analyticity and cluster
decomposition properties of the $S$--matrix to investigate the cusp structure.
In analogy to the corresponding $K^+$ decays discussed in the same reference,
an expansion in powers of the $\pi\pi$ scattering lengths $a$ is used as the essential
ordering principle, and the calculation is performed up to $O(a^2)$.
 In Ref.~\cite{gps}, in addition to analyticity and unitarity, chiral perturbation theory is used 
 for the evaluation of the real parts of the  $K\to 3\pi$ decay amplitudes at one loop.
In the present work, however, we rely on the non--relativistic effective 
field theory framework developed in Ref.~\cite{CGKR}.  
It is based on an effective Lagrangian, and as such satisfies all 
unitarity and analyticity constraints automatically.  
The coupling constants involved can be directly matched to $\pi\pi$ scattering
lengths, and the expansion in powers thereof as advocated in Ref.~\cite{Cabibbo:2005ez}
emerges naturally in a generalised power counting scheme.

Our presentation closely follows that of Ref.~\cite{CGKR}, allowing for a relatively
concise description of the procedure.  
We construct the most general non--relativistic Lagrangian required for
the process in question, and match the couplings to the $\pi\pi$ threshold parameters.
Thenceforth the calculation of the decay amplitude
up to two--loop order is straightforward.  
Our representation of tree, one--loop, and two--loop contributions
correctly reproduces the analytic structure with various branch points and cusps in the Mandelstam plane
throughout the physical region (and slightly beyond).
The pertinent calculation of the radiative corrections within the same framework
will follow in due course~\cite{radbfgkr}.

\begin{sloppypar}
\noindent{\bf 2.}
We consider the neutral and charged decay modes
$\kl(P_K)\to\pi^0(p_1)\pi^0(p_2)\pi^0(p_3)$ and 
$\kl(P_K)\to\pi^+(p_1)\pi^-(p_2)\pi^0(p_3)$. The kinematical variables
are defined as usual: $s_i=(P_K-p_i)^2$ with $p_i^2=M_i^2\, ,~i=1,2,3$, where
$M_{\pi^+}\doteq M_\pi$ and $M_{\pi^0}$ denote the masses of the charged and
neutral pions, respectively, and $\Delta_\pi=M_\pi^2-M_{\pi^0}^2\neq 0$. 
In the centre--of--mass  frame $P_K=(M_K,{\bf 0})$, with $M_K$ the
neutral kaon mass,
\be\label{eq:kin1}
p_i^0=\frac{M_K^2+M_i^2-s_i}{2M_K}\, ,\quad
{\bf p}_i^2=\frac{\lambda(M_K^2,M_i^2,s_i)}{4M_K^2}\, ,
\ee
where $\lambda(x,y,z)=x^2+y^2+z^2-2xy-2xz-2yz$ is the triangle 
function. Below we also use the velocities $v_{jk}$ and kinetic energies $T_i$,
\be\label{eq:kin2}
v_{jk}^2(s_i)=\frac{\lambda(s_i,M_j^2,M_k^2)}{s_i^2}\, ,\quad
T_i=p_i^0-M_i\,\,.
\ee
\end{sloppypar}
\noindent {\bf 3.}
We invoke the non--relativistic framework set up in Ref.~\cite{CGKR} for the evaluation of the pertinent 
decay amplitudes. In that framework,
the perturbative expansion is performed in terms of two formal parameters
$\epsilon$ and $a$. One counts the pion and kaon masses as $O(1)$, the pion momenta as
$O(\epsilon)$ and the pion mass difference $\Delta_\pi$ as
$O(\epsilon^2)$. In addition, each four--pion vertex is counted as a  quantity of order $a$.
 As these vertices are proportional to the $\pi\pi$ scattering lengths which are small,
one expects the expansion in $a$ to converge rapidly. We refer for a 
further discussion of the method to the original article~\cite{CGKR}. Here, we simply note that it is 
sufficient to provide the Lagrangian used -- the amplitudes then follow from a straightforward 
application of the rules provided in Ref.~\cite{CGKR}. 

\noindent{\bf 4.}
 The complete Lagrangian of the 
effective theory is ${\mathcal L}_K+{\mathcal L}_{\pi\pi}$, where ${\mathcal L}_K$ 
contains $\kl\to 3\pi$ vertices,  and ${\mathcal L}_{\pi\pi}$ describes elastic $\pi\pi$ scattering. 
In the following, we provide the Lagrangians necessary to calculate the amplitudes for $\kl\to 3\pi$ 
at order $\epsilon^4,a\epsilon^5, a^2\epsilon^2$.

\begin{sloppypar}
We start with the $\pi\pi$ interaction and consider the
following five physical channels in $\pi^a\pi^b\to\pi^c\pi^d$: $(ab;cd)=$
(1)~$(00;00)$, (2)~$(+0;+0)$, (3)~$(+-;00)$, (4)~$(+-;+-)$, (5)~$(++;++)$.
 [We omit the channel $\pi^-\pi^0\to\pi^-\pi^0$, because this amplitude is 
identical to $\pi^+\pi^0\to\pi^+\pi^0$ by charge invariance.] The Lagrangian takes the form
\end{sloppypar}

\vskip-5mm

\be\label{eq:l_pipi}
{\mathcal L}_{\pi\pi}=2\sum_\pm\Phi_\pm^\dagger W_\pm\bigl(i\partial_t-
W_\pm\bigr)\Phi_\pm 
+2\Phi_0^\dagger W_0\bigl(i\partial_t- W_0\bigr)\Phi_0
+\sum_{i=1}^5{\mathcal L}_{i}\, ,
\ee
where $\Phi_i$ is the non--relativistic pion field operator, 
$W_\pm=\sqrt{M_\pi^2-\triangle}$,  $ W_0=\sqrt{M_{\pi^0}^2-\triangle}$, with
$\triangle$ the Laplacian. Introducing further the notations
\newcommand{\Pp}{\mathcal{P}}
\eq\label{eq:Pn}
&&
(\Phi_n)_\mu=(\Pp_n)_\mu\Phi_n\, ,\quad
(\Phi_n)_{\mu\nu}=(\Pp_n)_\mu(\Pp_n)_\nu\Phi_n\, ,\quad
(\Pp_n)_\mu=(W_n,-i\nabla)\, ,
\nonumber\\[2mm]
&&
(\Phi_n^\dagger)_\mu=(\Pp^\dagger_n)_\mu\Phi^\dagger_n\, ,\quad
(\Phi_n^\dagger)_{\mu\nu}=
(\Pp^\dagger_n)_\mu(\Pp^\dagger_n)_\nu\Phi_n^\dagger\, ,\quad
(\Pp_n^\dagger)_\mu=(W_n,i\nabla)\, ,
\en
for $n=a,\,b,\,c,\,d$, one may write
\eq\label{eq:l_1-5}
{\mathcal L}_{i}&=&x_iC_i
\bigl(\Phi_c^\dagger\Phi_d^\dagger\Phi_a\Phi_b+h.c.\bigr)
\nonumber\\[2mm]
&+&x_iD_i\Bigl\{ (\Phi_c^\dagger)_\mu (\Phi_d^\dagger)^\mu\Phi_a\Phi_b
+\Phi_c^\dagger\Phi_d^\dagger (\Phi_a)_\mu (\Phi_b)^\mu
-h_i\Phi_c^\dagger\Phi_d^\dagger\Phi_a\Phi_b +h.c.\Bigr\}
\nonumber\\[2mm]
&+& \frac{u_iE_i}{2}
\Bigl\{\bigl(\Phi_c^\dagger (\Phi_d^\dagger)^\mu- 
(\Phi_c^\dagger)^\mu\Phi_d^\dagger \bigr)
\bigl((\Phi_a)_\mu\Phi_b-\Phi_a (\Phi_b)_\mu\bigr)
 + h.c.\Bigr\}
\nonumber\\[2mm]
&+&x_iF_i\biggl\{
(\Phi_c^\dagger)_{\mu\nu}(\Phi_d^\dagger)^{\mu\nu}\Phi_a\Phi_b
+\Phi_c^\dagger\Phi_d^\dagger(\Phi_a)_{\mu\nu}(\Phi_b)^{\mu\nu}
\nonumber\\[2mm]
&&\quad +\,2(\Phi_c^\dagger)_\mu (\Phi_d^\dagger)^\mu(\Phi_a)_\nu (\Phi_b)^\nu
+h_i^2 \Phi_c^\dagger\Phi_d^\dagger\Phi_a\Phi_b
\nonumber\\[2mm]
&&\quad -\,2h_i\bigl( (\Phi_c^\dagger)_\mu (\Phi_d^\dagger)^\mu\Phi_a\Phi_b
+\Phi_c^\dagger\Phi_d^\dagger (\Phi_a)_\mu (\Phi_b)^\mu\bigr)
+h.c.\biggr\}+\ldots~,
\en
with $h_i=s_i^t-\frac{1}{2}\,(M_a^2+M_b^2+M_c^2+M_d^2)$\,, where $s_i^t$ denotes
the physical threshold in the $i$th channel. Explicitly,
$h_1=2M_{\pi^0}^2$, $h_2=2M_\pi M_{\pi^0}$,
$h_3=3M_\pi^2-M_{\pi^0}^2$, $h_4=h_5=2M_\pi^2$. 
The ellipsis stands for terms of order $\epsilon^6$ in the $S$--wave and
for terms of order $\epsilon^4$ in the $P$-- and $D$--waves.
 The low--energy constants $C_i,D_i,E_i,F_i$ are matched to the physical threshold amplitudes below.
To simplify the resulting expressions, we have furthermore 
introduced the combinatorial factors $x_1=x_5=1/4$, $x_2=x_3=x_4=1$, $u_1=u_3=u_5=0$, 
$u_2=u_4=1$.
Finally,  we note that we omit local 6--pion couplings.
Their contribution to the $\kl\to 3\pi$ amplitude is purely imaginary in
the non--relativistic framework, and of order $\epsilon^4$.

\noindent{\bf 5.}
The couplings $C_i,D_i,E_i,F_i$ can be expressed in terms of
 the threshold parameters of the underlying relativistic theory.
 In the isospin symmetry 
limit, the expansion of the relativistic $\pi\pi$ scattering 
amplitude reads
\bea
{\rm Re}\,\bar T_i(s,t)&=&
\bar A_i\Bigl\{1+\frac{\bar r_i}{4M_\pi^2}\,\bigl(s-4M_\pi^2\bigr)
+\frac{\bar f_i}{16M_\pi^4}\,\bigl(s-4M_\pi\bigr)^2\Bigr\} 
\nnnl&&+\frac{3}{4}\bar A_i^P(t-u)+\ldots\fs
\eea 
The ellipsis stands for higher orders in $\epsilon$, 
e.g.  $D$--wave contributions. The bar indicates the isospin symmetric limit,
at $M_\pi=139.57$ MeV. In terms of the standard  scattering
lengths $a_0,a_2$ and $a_1$, one has
\bea\label{eq:isospin}
3\bar A_1&=&{N(a_0+2a_2)}{}\co\quad
2\bar A_2={Na_2}\co\quad
3\bar A_3={N(a_2-a_0)}\co\nnnl
6\bar A_4&=&{N(2a_0+a_2)}\co\quad
\bar A_5=Na_2\co \nnnl
2{\bar A}_2^P&=&N a_1\co\quad 2\bar A_4^P=N a_1\co\quad
\bar A_1^P=\bar A_3^P=\bar A_5^P=0\co\,\, N=32 \pi\,\,,
\eea
with $a_0=0.220\pm 0.005, a_2=-0.0444\pm 0.0010, a_0-a_2=0.265\pm 0.004, 
a_1=(0.379\pm 0.005)\times 10^{-1}M_\pi^{-2}$~\cite{Colangelo:2001df}.
The products $\bar A_i\bar r_i$ and $\bar A_i\bar f_i$
denote effective ranges and shape parameters, respectively.

Still in the isospin symmetry limit, the couplings $C_{i}$ are 
related to these threshold parameters according to %in the following manner. 
\be\label{eq:pipimatching1}
 2\bar C_i=\bar A_i\,\,\,,\quad 8 M_\pi^2 \bar D_i=\bar A_i\bar r_i\,
\,,\quad 32M_\pi^4\bar F_i=\bar A_i\bar f_i\,\,,\quad 
4\bar E_i=3\bar A_i^P\,,
\ee
where we have dropped higher--order terms in the threshold parameters.
 Taking isospin breaking into account, one 
finds at leading order in chiral perturbation theory~\cite{knechturech}
\be\label{eq:pipimatching2}
2C_{1,2,5}=\bar A_{1,2,5}(1-\eta),\,\,
2C_{3}=\bar A_3(1+\eta/3),\,\,
2C_{4}=\bar A_4(1+\eta),
\ee
where
$\eta=\Delta_\pi/M_{\pi}^2=6.5\times 10^{-2}$. 
 Isospin breaking in the remaining couplings
$D_i,E_i,F_i$ is expected to have a negligible effect on the analysis, 
and we propose to use for these couplings the
relations Eq.~\eqref{eq:pipimatching1} also in the real world, where isospin is
broken.

\begin{sloppypar}
\noindent{\bf 6.}
It remains to display the $\kl\rightarrow 3\pi$ 
Lagrangian,
\eq\label{eq:lag_K}
&&\hspace{-1cm}{\mathcal L}_K=2K^\dagger W_K\bigl(i\partial_t- W_K\bigr)K
 +L_0\bigl(K^\dagger\phin\phip\phim+h.c.\bigr)\nonumber\\[2mm]
&&\hspace{-1cm}+L_1\bigl(\kd(W_0-M_{\pin})\,\phin\phip\phim+h.c.\bigr) 
 +  L_2\bigl(\kd(W_0-M_{\pin})^2\,\phin\phip\phim+h.c.\bigr)\nonumber\\[2mm]
&&\hspace{-1cm}
+L_3\bigl(\kd\phin(W_\pm^2\phip\phim+\phip W_\pm ^2\phim-2W_\pm\phip W_\pm\phim)
        +h.c.\bigr)\nonumber\\[2mm]
&&\hspace{-1cm}
+\frac{1}{6}K_0\bigl(\kd\phin^3+h.c.\bigr)+\frac{1}{2}K_1\bigl(\kd\phin^2(W_0-M_{\pin})^2\phin
+h.c.\bigr)\,+\ldots\,,
\eea
where $K$ denotes the non--relativistic field for the $\kl$ meson, 
$ W_K=\sqrt{M_K^2-\triangle}$, and the ellipsis stands for the
higher--order terms in $\epsilon$. The couplings $L_i$, $K_i$ are
assumed to be real. Their contribution to the  decay matrix elements 
 at tree--level is  provided below.
\end{sloppypar}

 The tree--level expressions
 for the amplitudes, generated by ${\mathcal L}_K$, are
 modified by final state interactions of the pions, generated by  loops
 evaluated with ${\mathcal L}_{\pi\pi}$. We use the notation
\bea\label{eq:defexpand}
{\mathcal M}_{000}={\mathcal M}_N^{\words{tree}}
                  +{\mathcal M}_N^{\words{1-loop}}
                  +{\mathcal M}_N^{\words{2-loops}}+\ldots\,\,\,\,
[\kl\to\pi^0\pi^0\pi^0]\,\,,\nonumber\\[1mm]
{\mathcal M}_{+-0}={\mathcal M}_C^{\words{tree}}
                  +{\mathcal M}_C^{\words{1-loop}}
                  +{\mathcal M}_C^{\words{2-loops}}+\ldots\,\,\,\,
[\kl\to\pi^+\pi^-\pi^0]\,\,
\eea
for the decay amplitudes and the Condon--Shortley phase convention for 
the pions. Our amplitudes are
normalised such that the decay rates are given by
\be
d\Gamma=\frac{1}{2M_K}(2\pi)^4\delta^{(4)}(P_f-P_i)
{|\mathcal M|}^2\prod_{i=1}^3
\frac{d^3{\bf p}_i}{2(2\pi)^3p_i^0}\,.
\ee
In the case of $\kl\to 3\pi^0$, the right hand side must be divided by 3!=6.

\noindent{\bf 7.}
The {\it tree} amplitudes are
\newcommand{\xone}{X_{1}}
\newcommand{\xtwo}{X_{2}}
\newcommand{\xthree}{X_{3}}
\bea\label{eq:tree1}
\mathcal{ M}_0^{\words{tree}} &=& K_0+K_1\left(\xone^2+\xtwo^2+\xthree^2\right) \,,\nn
\mathcal{ M}_{\pm}^{\words{tree}} &=& L_0+L_1\xthree+L_2\xthree^2+L_3(\xone-\xtwo)^2 \, ,
\eea
where $X_i = p_i^0-\mn\,$.
This representation is equivalent to
\bea\label{eq:tree2}
\mathcal{ M}_0^{\words{tree}} &=& U_0+U_1\Big(u^2+\frac{v^2}{3}\Big), \nn
\mathcal{ M}_{\pm}^{\words{tree}} &=& V_0+V_1(s_3-s_c)+V_2(s_3-s_c)^2+V_3(s_2-s_1)^2,
\eea
 where
\begin{align}\label{eq:tree3}
u &= s_3-s_n ~, &v &= s_2-s_1 ~,\nn
s_n &= \frac{M_K^2+3\mn^2}{3} ~, &s_c &= \frac{M_K^2+M_{\pi^0}^2 + 2M_\pi^2}{3} ~.
%p_i^0 &= \frac{M_K^2+M_i^2-s_i}{2M_K}.
\end{align}
The relations between the coefficients $U_i,V_i$ and $L_i,K_i$ are displayed in Appendix A.

\noindent{\bf 8.}
The {\it one--loop} contributions are proportional to the basic integral
\begin{equation}\label{eq:defJ}
\hspace*{-4.mm}J_{ab}(P^2)=\int\frac{d^Dl}{i(2\pi)^D}\,
\frac{1}{2w_a({\bf l})2w_b({\bf P}-{\bf l})}\,
\frac{1}{(w_a({\bf l})-l_0)(w_b({\bf P}-{\bf l})-P_0+l_0)}\, ,
\end{equation}
with 
 $w_\pm({\bf p})=\sqrt{M_\pi^2+{\bf p}^2}$,
$w_0({\bf p})=\sqrt{M_{\pi^0}^2+{\bf p}^2}$ and $P^2=P_0^2-{\bf P}^2$. In the limit $D\to 4$,
\vspace*{-.1cm}
\begin{equation}\label{eq:functionJ}
J_{ab}(P^2)=\frac{i}{ 16\pi} v_{ab}(P^2) \,\, ,
\end{equation}
which is a quantity of order $\epsilon$.
 In order to make the formulae more transparent,
 we modify the notation for the couplings $C_i,D_i,E_i,F_i$ ,
\be
(C_1,C_2,C_3,C_4,C_5)=(C_{00},C_{+0},C_{x},C_{+-},C_{++})\,,
\ee
and analogously for the $D_i, E_i,F_i$.
In the following, we use $J_{-0}=J_{+0}$ throughout,  and denote the couplings for 
$\pi^-\pi^0\to\pi^-\pi^0$ with  index $+$ as well, $C_{-0}=C_{+0}$, etc.
 We then find
\newcommand{\nnnlt}{\nonumber\\[2mm]}
\newcommand{\nnnlv}{\nonumber\\[4mm]}
\newcommand{\yoz}{Y_{1n}}
\newcommand{\ytz}{Y_{3n}}
\newcommand{\zoz}{Z_{1}}
\newcommand{\ztz}{Z_{3}}

\newcommand{\yoc}{Y_{1c}}
\newcommand{\ytc}{Y_{3c}}

\newcommand{\yonc}{Y_{1nc}}

\newcommand{\wm}{Z_1^-}
\newcommand{\nwp}{Z_1^+}

\bea\label{eq:1loop}
\mathcal{ M}_0^{\words{1--loop}} &=&
\Big\{ B_{0}^{(1)}(s_1)J_{00}(s_1)+ (s_1 \leftrightarrow
s_2 )+ (s_1 \leftrightarrow s_3) \Big\}\nnnlt
&+& \Big\{B_{0}^{(2)}(s_1)J_{+-}(s_1) +(s_1 \leftrightarrow
s_2 )+ (s_1 \leftrightarrow s_3) \Big\},\nnnlt
\mathcal{M}_{\pm}^{\words{1--loop}} &=& 
B_{\pm}^{(1)}(s_3)J_{00}(s_3)+B_{\pm}^{(2)}(s_3)J_{+-}(s_3)\nnnlt
&+&\Big\{ B_{\pm}^{(3)}(s_1,s_2,s_3)J_{+0}(s_1)+(s_1 \leftrightarrow s_2) \Big\},
\eea
with
\bea
B_{0}^{(1)}(s_1) &=& \left(C_{00}+D_{00}\yoz+F_{00}\yoz^2 \right)
 \left \{ K_0 + K_1 \Bigl[\xone^2 +2\zoz^2+\frac{{\bf Q}^2_1}{6s_1}\yoz \Bigr] \right\},\nnnlv
B_{0}^{(2)}(s_1) &=&
2\left(C_x+D_x\yoc+F_x \yoc^2 \right)\left\{L_0 
+ L_1\xone+L_2\xone^2 +L_3 \frac{{\bf Q}_1^2  }{3s_1}\yoc \right\}\,,\nnnlv
B^{(1)}_{\pm}(s_3) &=& \left(C_x+D_x \ytc + F_x \ytc^2 \right) \left\{ K_0
+ K_1 \Bigl[\xthree^2
+2\ztz^2+\frac{{\bf Q}^2_3}{6s_3}\ytz \Bigr] \right\},\nnnlv
B^{(2)}_{\pm}(s_3) &=& 2\left(C_{+-} \!+\! D_{+-} \ytc \!+\! F_{+-} \ytc^2 \right) \left\{\!L_0+L_1\xthree
+L_2 \xthree^2
+L_3 \frac{{\bf Q}_3^2}{3s_3} \ytc
\right\},\nonumber\eea

\vskip-9mm

\bea
B^{(3)}_{\pm}(s_1,s_2,s_3) &=& 2\left(C_{+0}+D_{+0} \yonc + F_{+0} \yonc^2 \right)
\biggl\{ L_0+L_1 \wm  \nnnlt
&+&
 L_2 \Bigl[(\wm)^2+\frac{{\bf Q}^2_1 \, q_{23}^2(s_1)}{3 s_1} \Bigr]
+\! L_3 \Bigl[(\nwp \!-\! X_1)^2+\frac{{\bf Q}^2_1 \,q_{23}^2(s_1)}{3 s_1} \Bigr] 
\biggr\} \nnnlv
&-& \frac{1}{3}E_{+0}\frac{q_{23}^2(s_1)}{s_1 M_K}
\Big(\dpi(\mc^2-M_K^2)+s_1(s_3-s_2) \Big)\nnnlv
&&\times \biggl\{ L_1 +2L_2 \wm
+2L_3 \left[X_1-\nwp\right] \biggr\}
+O(\dpi^2)\,.
\eea
We have used the abbreviations
\eq
Q_1^0&=&p_2^0+p_3^0\,\,\,(\words{cycl.})\co\qquad
{\bf Q}_i^2=\frac{\lambda(M_K^2,M_i^2,s_i)}{4M_K^2},\nn
q^2_{lm}(s_k) &=& \frac{\lambda(s_k,M_l^2,M_m^2)}{4s_k}\qquad ( k\neq l \neq
m \neq k) \co\nn
Y_{in}&=&s_i-4\mn^2\co\,Y_{ic}=s_i-4\mc^2\co\, Y_{inc}=s_i-(\mn+\mc)^2\co\, \nn
Z_i&=&\frac{Q_i^0}{2}-\mn \co \, 
Z_i^\pm = \frac{Q_i^0}{2}\left(1\pm\frac{\Delta_\pi}{s_i}\right)-\mn\fs
\en

\begin{figure}[t]
\begin{center}
\includegraphics[width=9.cm]{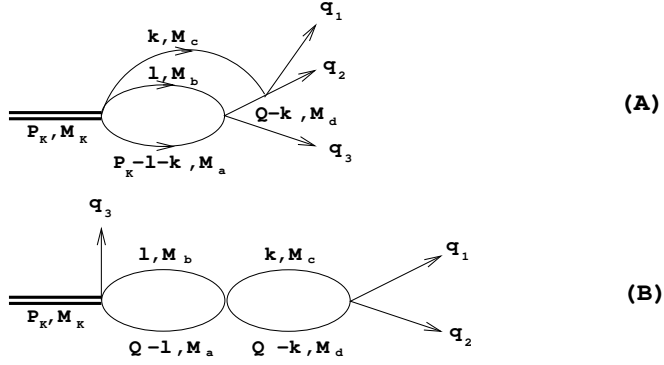}
\end{center}
\caption{Two topologically distinct non--relativistic two--loop graphs 
describing the final--state $\pi\pi$ rescattering in the decay $K\to 3\pi$, with
$Q^\mu=(q_1+q_2)^\mu$.}
\label{fig:2loop_text}
\end{figure}

\noindent {\bf 9.}
There are two topologically distinct two--loop graphs that describe
pion--pion rescattering in the final state, see Fig.~\ref{fig:2loop_text}.
 At the order of accuracy we are working, it is sufficient to consider the case
of non--derivative couplings. In this case, the contributions of 
both diagrams depend only on the variable $s$, where
\begin{equation}
Q^\mu=(q_1+q_2)^\mu \,\,,\,\,Q^2=s\,\, .
\end{equation}
The diagram
in  Fig.~\ref{fig:2loop_text}B, apart from a factor 
containing coupling constants, is given by a product of two one--loop
diagrams which were already calculated in Eq.~\eqref{eq:functionJ}.
The non--trivial contribution from Fig.~\ref{fig:2loop_text}A
is proportional to
\eq\label{eq:twoloop_F}
{\mathcal M}(s)
&=&\int\frac{d^D l}{i(2\pi)^D}\,\frac{d^D k}{i(2\pi)^D}\,
\nonumber\\[1mm]
&\times&
\frac{1}{2w_a({\bf l}+{\bf k})}\,
\frac{1}{w_a({\bf l}+{\bf k})-M_K+l^0+k^0}\,
\frac{1}{2w_b({\bf l})}\, 
\frac{1}{w_b({\bf l})-l^0}\,
\nonumber\\[1mm]
&\times&\frac{1}{2w_c({\bf k})}\, 
\frac{1}{w_c({\bf k})-k^0}\,
\frac{1}{2w_d({\bf Q}-{\bf k})}\, 
\frac{1}{w_d({\bf Q}-{\bf k})-Q^0+k^0}\, .
\en
A short discussion of this integral is given  in Ref.~\cite{CGKR}. There, it is shown that one may write
\be\label{eq:mostgeneral}
{\mathcal M}(s)
=F\bigl(M_a,M_b,M_c,M_d;s\bigr)+\ldots ,
\ee
where $F$ is ultraviolet finite and
contains the full non--analytic behaviour of the two--loop diagram
in the low--energy domain, whereas the ellipsis denotes terms that amount to a redefinition 
of the tree--level couplings in ${\mathcal L}_K$ and which are therefore dropped.
 A one--dimensional  integral representation for $F$  is provided in Ref.~\cite{CGKR}. The relevant 
integrals can be performed analytically -- the result is displayed in  Appendix B.
\begin{sloppypar}
Below, we use the notation $F_i(\ldots;s)$ for the integral $F(\ldots;s)$, 
evaluated at ${\bf
    Q}^2=\lambda(M_K^2,M_{\pi^i}^2,s)/4M_K^2$, with $i=\pm,0$.
\end{sloppypar}

\begin{figure}[t]
\begin{center}
\includegraphics[width=\linewidth]{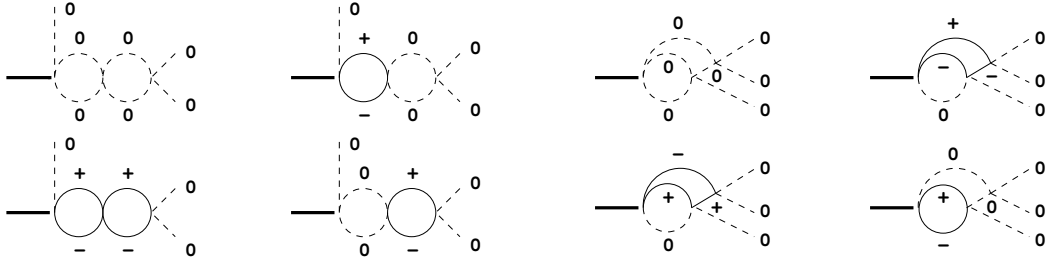}
\end{center}
\caption{Two--loop graphs contributing to the decay $\kl\to\pi^0\pi^0\pi^0$
in the non--relativistic effective theory. The graphs obtained by a 
permutation of  identical particles in the final state are not
 shown.} \vskip 2mm
\label{fig:neutral}
\end{figure}

\begin{figure}[t]
\begin{center}
\includegraphics[width=\linewidth]{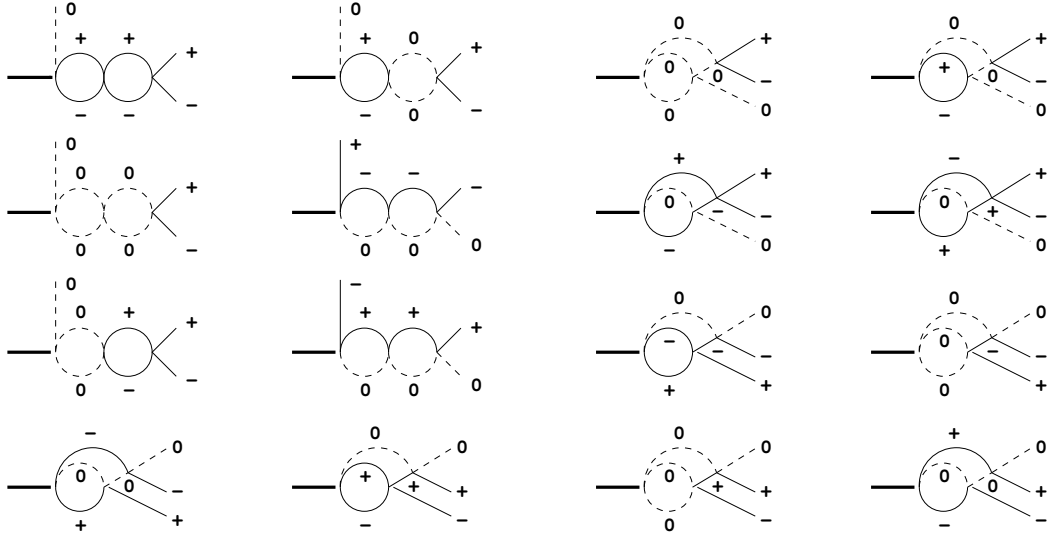}
\end{center}
\caption{Two--loop graphs contributing to the decay $\kl\to\pi^+\pi^-\pi^0$
in the non--relativistic effective theory.}
\label{fig:charged}
\end{figure}

Evaluating the diagrams displayed in Figs.~\ref{fig:neutral} and \ref{fig:charged}, 
we find for the amplitudes at order $a^2\epsilon^2$
\bea\label{eq:2loop1}
\mathcal{M}_0^{\words{2-loops}} &=&
\Big\{\mathcal{M}^A_0(s_1)+\mathcal{M}^B_0(s_1) +(s_1
\leftrightarrow s_2 ) + (s_1 \leftrightarrow s_3  ) \Big\},\nn
\mathcal{M}_{\pm}^{\words{2-loops}} &=&
\mathcal{M}^A_{\pm}(s_1,s_2,s_3)+\mathcal{M}^B_{\pm}(s_1,s_2,s_3),
\eea

\vskip-2mm

where

\vskip-2mm

\bea\label{eq:2loop2}
\mathcal{M}^A_0(s_1) &=& 2\, C_{00}^2\, K_0\, F_{0}(\mn,\mn,\mn,\mn;s_1)\nn
&+& 8\, C_{+0}\, C_x\, L_0\, F_{0}(\mn,\mc,\mc,\mc;s_1)\nn
&+& 4\, C_{00}\,C_x\, L_0\, F_{0}(\mc,\mc,\mn,\mn;s_1),\nonumber\\[3mm]
\mathcal{M}^B_0(s_1) &=& C_{00}^2\, K_0\, J^2_{00}(s_1) +  4\,
C_x\,C_{+-}\, L_0\, J^2_{+-}(s_1)\nn
&+&\big(2\, C_x^2\, K_0 + 2\, C_{00}\, C_x\, L_0\big)J_{+-}(s_1) J_{00}(s_1),
\eea

\vskip-2mm

and

\vskip-2mm

\bea\label{eq:2loop3}
\mathcal{M}^A_{\pm}(s_1,s_2,s_3) &=& 2\, C_{00}\,C_x\, K_0\,
F_{0}(\mn,\mn,\mn,\mn;s_3)\nn
&+& 4\, C_x^2\,L_0\, F_{0}(\mc,\mc,\mn,\mn;s_3)\nn
&+& 8\, C_{+0}\,C_{+-}\, L_0\, F_{0}(\mn,\mc,\mc,\mc;s_3)\nn
&+& \Big\{ 4\, C_{+0}\, C_{+-}\,L_0\, F_+(\mc,\mc,\mn,\mc;s_1) \nn
&+&2\,C_{+0}\,C_x\, K_0\, F_+(\mn,\mn,\mn,\mc;s_1)\nn
&+& 4\, C_{+0}^2\, L_0\, F_+(\mc,\mn,\mc,\mn;s_1) 
+ \big(s_1 \leftrightarrow s_2\big)  \Big\},\nonumber\\[3mm]
\mathcal{M}^B_{\pm}(s_1,s_2,s_3) &=& 4\, C_{+-}^2\, L_0\, J^2_{+-}(s_3) +
  C_{00}\,C_x\, K_0\, J^2_{00}(s_3)  \nn
&+& \big( 2\, C_x\, C_{+-}\,K_0 + 2\, C_x^2\, L_0 \big)J_{+-}(s_3) J_{00}(s_3)\nn
&+& \Big\{4\, C_{+0}^2\,L_0\, J^2_{+0}(s_1) +\big(s_1 \leftrightarrow s_2\big)  \Big\}.
\eea

\noindent{\bf 10.}
The decay amplitudes depend on the six {\em real} $\kl\to 3\pi$ coupling
constants $L_i,K_i$ and on the threshold parameters for $\pi\pi$
scattering. Combining the tree-- and one--loop result
Eqs.~\eqref{eq:tree1}, \eqref{eq:1loop} with the two--loop contributions
Eqs.~\eqref{eq:2loop1}, we obtain the neutral and charged decay
amplitudes up to and  
including terms of order $\epsilon^4,a\epsilon^5$ and $a^2\epsilon^2$,
expressed in terms of the one-- and two--loop integrals $J$ and $F$ displayed
in Eqs.~\eqref{eq:functionJ} and \eqref{eq:functionF1}--\eqref{eq:functionF3}, respectively.
 [We have dropped some of the contributions at order $\epsilon\Delta_\pi^2$. In particular, 
$D$--waves generate contributions of this type. We expect them to be completely 
negligible.] This
representation is valid in the whole decay region, and is the main result
of this article.

The decay amplitude $\kl \to\pi^0\pi^0\pi^0$ obeys what we refer to as the
{\it threshold theorem}: the coefficient of the leading non--analytic piece,
which is proportional to $v_{+-}(s_3)$, is given by a product of two
factors, the decay amplitude $\kl\to\pi^0\pi^+\pi^-$ and the scattering
amplitude $\pi^+\pi^-\to\pi^0\pi^0$, both evaluated at threshold~\cite{Cabibbo:2004gq}.
 Of course, aside from the determination of the leading term in $v_{+-}$,
our approach also allows a systematic evaluation of higher--order
contributions $v_{+-}^3,v_{+-}^5\ldots$\,.

\noindent {\bf 11.}
We now compare the content of this letter with the  work of Cabibbo
and Isidori~\cite{Cabibbo:2005ez} (CI), who  use an alternative
method to construct the $K\to 3\pi$ decay amplitudes. 
Conceptual aspects of the  two methods  were already  
discussed in Ref.~\cite{CGKR} for the case of the charged kaon decays $K^+\to 3\pi$. 
 In particular, it was pointed out that
the amplitudes agree at order $a$, whereas they differ at order $a^2$ away from threshold,
because the method used by CI does not reproduce the correct analytic properties 
of the amplitudes at two--loop order. [On the other hand, the two amplitudes lead to
very similar results for the scattering lengths when fitted to $K^+\to 3\pi$ 
data~\cite{dilellakaon07}.]
Analogous comments apply in the case of $\kl\to3\pi$ considered here.
  Comparing the  expressions in detail, we note that the final result Eqs.~(4.61)--(4.67) in CI
 does contain some (but not all) of the terms evaluated above. In this sense, 
the expansion of  the decay amplitudes  presented here is more systematic and complete.
As to the terms  retained in CI, we note that,
 aside from obvious typos, we do agree in $\kl\to 3\pi^0$ at order $a$ in the physical region, 
and at order $a^2$ at  the thresholds $s_i=4\mc^2$. 
 In the charged channel $\kl\to \pi^+\pi^-\pi^0$, a graph is omitted in 
CI. It contributes  at order $a\epsilon$ and generates a cusp
 at the edge of physical phase space. 

\pagebreak[4]

\noindent {\bf 12.}
We add a remark concerning $\eta\to 3\pi^0$ and $\eta\to \pi^+\pi^-\pi^0$ decays.
These processes can be analysed in a completely 
analogous fashion.
 Indeed, the $\pi\pi$ scattering 
amplitudes remain the same, whereas the polynomial Lagrangian for $\eta\to 3\pi$ can be
 obtained from the $K\to 3\pi$ one by replacing 
field operators and particle masses in the Lagrangian Eq.~\eqref{eq:lag_K}, 
($K,M_K$) $\to$ ($\eta,M_\eta$).
The tree amplitudes analogous to Eq.~\eqref{eq:tree1} become
\bea\label{eq:etatree}
\mathcal{ M}_{0}^{\eta\words{tree}} &=& K^\eta_0+
K^\eta_1\left(\xone^2+\xtwo^2+\xthree^2\right) \,,\nn
\mathcal{ M}_{\pm}^{\eta\words{tree}} &=& L^\eta_0+L^\eta_1\xthree+
L^\eta_2\xthree^2+L^\eta_3(\xone-\xtwo)^2 \, ,
\eea
with $X_i=p_i^0-M_{\pi^0}$, and  with obvious notation otherwise. The relation to an 
alternative expansion in the conventional $\eta\to3\pi$ Dalitz plot 
variables is provided in Appendix A.
 Furthermore, the one-- and two--loop results in Eqs.~\eqref{eq:1loop}, \eqref{eq:2loop1} 
 can simply be taken over,
with the replacements $(K_i,L_i,M_K)\to(K_i^\eta,L_i^\eta,M_\eta)$ everywhere.
 Because $\Gamma_{K_L\to\pi^+\pi^-\pi^0}/\Gamma_{K_L\to3\pi^0}$ $\sim$ 
  $\Gamma_{\eta\to\pi^+\pi^-\pi^0}/\Gamma_{\eta\to3\pi^0}$, 
we expect that the 
strength of the cusp effect  in the neutral channel $\eta\to 3\pi^0$ is
of the same order as the one in $\kl\to 3\pi^0$, i.e., 
much less visible  than in the charged channel $K^+\to \pi^+\pi^0\pi^0$.

\begin{sloppypar}
\noindent {\bf 13.}
In summary, we have investigated $\kl\to 3\pi$ decays within a
non--relativistic effective Lagrangian framework.  The amplitudes are
calculated in a systematic double expansion in the pion momenta
 (counted as quantities of order $\epsilon$), and in the
threshold parameters of elastic $\pi\pi$ scattering (generically
denoted by $a$).  We provide an explicit representation of the amplitudes
at order $\epsilon^4, a\epsilon^5, a^2\epsilon^2$.
The representation is valid in the 
physical decay region, and contains the six (real) $\kl\to 3\pi$ coupling constants
$L_i,K_i$ and  the threshold parameters $a$.  The very same amplitude can be used, with
trivial modifications described above, for a cusp analysis in $\eta\to 3\pi$.
\end{sloppypar}

Our amplitudes  differ from  the ones of Cabibbo and
Isidori \cite{Cabibbo:2005ez} when compared in detail -- in particular, we do retain  all terms 
at the above
 mentioned order in the low--energy expansion.
 For this reason, we believe that it is important to check whether our
expressions for the amplitudes lead to scattering lengths that are in agreement with the ones
generated with the amplitudes presented in Ref.~\cite{Cabibbo:2005ez}.  

It remains to investigate  radiative corrections, which can be evaluated
in the field--theoretical framework used here in a standard manner. 
The effects generated by the $\pi^+\pi^-$ bound state at the $\pi^+\pi^-$
threshold can also be investigated within
the same approach  
\cite{nrqft6,nrqft7,nrqft8,nrqft9,nrqft10,nrqft11,nrqft12,nrqft13,nrqft14}, see also Ref.~\cite{tarasovk3pi}.
We plan to include these effects in forthcoming publications~\cite{radbfgkr}. For the evaluation of 
radiative corrections in $K^+\to \pi^+\pi^+\pi^-$ in the framework of scalar QED, 
we refer the reader to the recent interesting article by Isidori~\cite{isidorirad}.

\pagebreak[4]

\begin{sloppypar}
{\it Acknowledgements.}
We thank Gilberto Colangelo,  Gino Isidori and Heiri Leutwyler for useful comments 
on the manuscript. Partial financial support under the EU Integrated Infrastructure
Initiative Hadron Physics Project (contract number RII3--CT--2004--506078)
and DFG (SFB/TR 16, ``Subnuclear Structure of Matter'') is gratefully
acknowledged. This work was  supported  by the Swiss
National Science Foundation,  and by EU MRTN--CT--2006--035482
(FLAVIA{\it net}). 
 One of us (J.G.) is grateful to the Alexander von Humboldt--Stiftung and to 
the Helmholtz--Gemeinschaft for the award of  a prize 
 that allowed him to stay at the HISKP at the University of Bonn, 
where part of this work was performed. 
He also thanks the HISKP for the warm hospitality during these stays.
\end{sloppypar}

\setcounter{equation}{0}
\renewcommand{\theequation}{A.\arabic{equation}}

{\bf Appendix A}

The coefficients $U_i,V_i$ are given by
\bea
U_0 &=& K_0+\frac{3 K_1}{4 M_K^2}\big((M_K-\mn)^2-s_n \big)^2 ~,\qquad U_1 = \frac{3K_1}{8 M_K^2}~,\\
V_0 &=& L_0 +\frac{L_1}{2 M_K} \big((M_K-\mn)^2-s_c
\big)+\frac{L_2}{4 M_K^2}\big((M_K-\mn)^2-s_c \big)^2~,\nn
V_1 &=& \frac{L_2}{2 M_K^2} \big(s_c-(M_K-\mn)^2 \big)-\frac{L_1}{2 M_K}~, \quad
V_2 = \frac{L_2}{4 M_K^2}~, \quad V_3 = \frac{L_3}{4 M_K^2}~. \nonumber
\eea
The inverse relations read
\bea
K_0 &=& U_0-2U_1\big((M_K-\mn)^2-s_n\big)^2 ~, \quad
K_1 = \frac{8}{3}M_K^2 U_1 ~,\nn
L_0 &=& V_0+V_1 \big((M_K-\mn)^2-s_c\big)+V_2 \big((M_K-\mn)^2-s_c\big)^2~,\nn
L_1 &=& 4M_K V_2\big(s_c-(M_K-\mn)^2\big)-2M_KV_1 ~, \nn
L_2 &=& 4 M_K^2V_2 ~,\quad L_3 = 4M_K^2V_3 ~.
\eea
For the decays $\eta\to 3\pi^0$ and $\eta\to\pi^+\pi^-\pi^0$,  
the tree amplitudes Eq.~\eqref{eq:etatree} may be written in the alternative expansion
\be
\mathcal{ M}_{0}^{\eta\words{tree}} = u_0+ u_1 z ~, \qquad
\mathcal{ M}_{\pm}^{\eta\words{tree}} = v_0+v_1y+v_2y^2+v_3x^2 ~\co
\ee
with the conventional $\eta\to3\pi$ Dalitz plot variables 
\be
x= \frac{\sqrt{3}(p_1^0-p_2^0)}{Q_\eta} ~, \quad
y= \frac{3(p_3^0-\mn)}{Q_\eta}-1 ~, \quad
z = \frac{2}{3}\sum_{i=1}^3\Bigl(\frac{3p_i^0-M_\eta}{Q_{\eta 0}}\Bigr)^2 ~,
\ee
where $Q_\eta=M_\eta-2\mc-\mn$, $Q_{\eta 0}=M_\eta-3\mn$.
The coefficients of the two representations are related by
\bea
u_0 &=& K^\eta_0+\frac{Q_{\eta 0}^2}{3}K^\eta_1 ~, \quad
u_1  =  \frac{Q_{\eta 0}^2}{6}K^\eta_1 ~,\quad
v_0  =  L^\eta_0+\frac{Q_\eta}{3}L^\eta_1+\frac{Q_\eta^2}{9}L^\eta_2 ~, \nn
v_1 &=& \frac{Q_\eta}{3}\Bigl(L^\eta_1+\frac{2}{3}Q_\eta L^\eta_2\Bigr) ~, \quad
v_2  =  \frac{Q_\eta^2}{9}L^\eta_2 ~, \quad
v_3  =  \frac{Q_\eta^2}{3}L^\eta_3 ~,
\eea
or reversely by
\bea
K_0^\eta &=& u_0-2u_1 ~, \quad
K_1^\eta = \frac{6}{Q_{\eta 0}^2} u_1 ~, \quad
L_0^\eta = v_0 - v_1 + v_2 ~, \nn
L_1^\eta &=& \frac{3(v_1-2v_2)}{Q_\eta} ~, \quad
L_2^\eta  =  \frac{9v_2}{Q_\eta^2} ~, \quad 
L_3^\eta =  \frac{3v_3}{Q_\eta^2} ~.
\eea

\setcounter{equation}{0}
\renewcommand{\theequation}{B.\arabic{equation}}

{\bf {Appendix B}}

The analytic expression for the two--loop function $F$ reads
\begin{equation}
F(M_a,M_b,M_c,M_d,s) ~=~ {\mathcal N}\,(2A\, f_1+B\, f_0) + O(\epsilon^4) \, ,
\label{eq:functionF1}\end{equation}
with
\bea
{\mathcal N}&=&\frac{1}{256\pi^3\sqrt{s}}\, 
\biggl(1-\frac{2(M_a^2+M_b^2)}{s_0}+\frac{(M_a^2-M_b^2)^2}{s_0^2}\biggr)^{1/2}
\frac{1}{\sqrt{\Delta^2-\frac{(1+\delta)^2}{4}\,{\bf Q}^2}}\, ,
\nonumber\\[2mm]
f_0&=&4\bigl(v_1+v_2-\bar v_{2}+h\bigr)\, ,\nnnl
f_1&=&
\frac{4}{3}\,\bigl(y_1(v_1-1)+y_2(v_2-1)-\bar y_{2}(\bar v_{2}-1) +h \bigr)\, ,
\nonumber\\[2mm]
h&=&\frac{1}{2}\ln\biggl(\frac{1+{\bf Q}^2/s}{1+\bar{\bf Q}^2/\bar s}\biggr) \, ,
\quad \bar {\bf Q}^2={\bf Q}^2(\bar s)\,,
\nonumber\\[2mm]
v_i&=&\sqrt{-y_i}\,\arctan\frac{1}{\sqrt{-y_i}} \, ,\quad i=1,2 ~;\quad
\bar v_{2}=\sqrt{-\bar y_{2}}\,\arctan\frac{1}{\sqrt{-\bar y_{2}}}\, , \quad
\nnnl
y_{1,2} &=& \frac{-B \mp \sqrt{B^2-4AC}}{2A}\, ,\quad \bar y_{2}=y_2(\bar s)\,  \nnnl
\nonumber\\[2mm]
A&=& -\frac{{\bf Q}^2}{s}\,(M_c^2+\Delta^2)\, ,\quad
B=q_0^2-\Delta^2+\frac{{\bf Q}^2}{s}\, M_c^2\, ,\quad
C=-q_0^2\, ,\nnnl
s_0&=&M_K^2+M_c^2-2M_K\biggl( M_c^2+\frac{{\bf Q}^2(1+\delta)^2}{4}\biggr)^{1/2}\, ,
\nonumber\\[1mm]
q_0^2&=&\frac{\lambda(s,M_c^2,M_d^2)}{4s}\, ,\quad\quad 
\bar s=(M_c+M_d)^2\, ,
\nonumber\\[1mm]
\Delta^2&=&\frac{\lambda(M_K^2,M_c^2,(M_a+M_b)^2)}{4M_K^2}\, ,\quad\quad
\delta=\frac{M_c^2-M_d^2}{s}\,.\label{eq:functionF2}
\eea
The $\arctan$ is understood to be evaluated according to 
\begin{equation}\label{eq:functionF3}
\arctan x = \frac{1}{2i}\ln\frac{1+ix}{1-ix} ~,
\end{equation}
and $s$ is given a small positive imaginary part in all  arguments, $s \to s+i\epsilon$.

The analytic formula Eq.~\eqref{eq:functionF1} is exact at $O(\epsilon^2)$, 
and thus at the order considered in Ref.~\cite{CGKR}. 
It differs by a few percent from the integral representation given in Ref.~\cite{CGKR}.

\ed
\begin{thebibliography}{99}

\bibitem{Cabibbo:2004gq}
  N.~Cabibbo,
  Phys.\ Rev.\ Lett.\  {\bf 93} (2004) 121801
  [arXiv:hep-ph/0405001].
  %%CITATION = HEP-PH 0405001;%%

\bibitem{Cabibbo:2005ez}
  N.~Cabibbo and G.~Isidori,
  JHEP {\bf 0503} (2005) 021
  [arXiv:hep-ph/0502130].
  %%CITATION = HEP-PH 0502130;%%

\bibitem{CGKR}
  G.~Colangelo, J.~Gasser, B.~Kubis and A.~Rusetsky,
  %``Cusps in K --> 3pi decays,''
  Phys.\ Lett.\  B {\bf 638} (2006) 187
  [arXiv:hep-ph/0604084].
  %%CITATION = PHLTA,B638,187;%%

\bibitem{Batley}
 J.~R.~Batley {\it et al.}  [NA48/2 Collaboration],
  %``Observation of a cusp-like structure in the pi0 pi0 invariant mass
  %distribution 
  %from K+- $\to$ pi+- pi0 pi0 decay and determination of the pi pi
  %scattering lengths,''
  Phys.\ Lett.\ B {\bf 633} (2006) 173
  [arXiv:hep-ex/0511056].
  %%CITATION = HEP-EX 0511056;%%

\bibitem{dilellakaon07}
  L.~Di Lella: {\it Review of $\pi\pi$ scattering measurements in $K$ decays}, 
talk given at: Kaon'07,
  May 21--25, 2007, Frascati, Italy, to appear in the proceedings.


\bibitem{isidorirad}
  G.~Isidori,
  %``Soft-photon corrections in multi-body meson decays,''
  arXiv:0709.2439 [hep-ph].
  %%CITATION = ARXIV:0709.2439;%%

\bibitem{Pislak:2003sv}
  S.~Pislak {\it et al.},
  %``High statistics measurement of K(e4) decay properties,''
  Phys.\ Rev.\ D {\bf 67} (2003) 072004
  [arXiv:hep-ex/0301040].
  %%CITATION = HEP-EX 0301040;%%


\bibitem{blochkaon07}
B.~Bloch-Devaux,
 {\it Recent results from NA48/2 on Ke4 decays and interpretation in term of $\pi\pi$
 scattering lengths},  talk given at: Kaon'07,
  May 21--25, 2007, Frascati, Italy, to appear in the proceedings.
 


\bibitem{Adeva:2005pg}
  B.~Adeva {\it et al.}  [DIRAC Collaboration],
  %``First measurement of the pi+ pi- atom lifetime,''
  Phys.\ Lett.\ B {\bf 619} (2005) 50
  [arXiv:hep-ex/0504044].
  %%CITATION = HEP-EX 0504044;%%

\bibitem{Colangelo:2000jc}
  G.~Colangelo, J.~Gasser and H.~Leutwyler,
  %``The pi pi S-wave scattering lengths,''
  Phys.\ Lett.\ B {\bf 488} (2000) 261
  [arXiv:hep-ph/0007112].
  %%CITATION = HEP-PH 0007112;%%

\bibitem{Colangelo:2001df}
  G.~Colangelo, J.~Gasser and H.~Leutwyler,
  %``pi pi scattering,''
  Nucl.\ Phys.\ B {\bf 603} (2001) 125
  [arXiv:hep-ph/0103088].
  %%CITATION = HEP-PH 0103088;%%


\bibitem{descotes}
 S.~Descotes-Genon, N.~H.~Fuchs, L.~Girlanda and J.~Stern,
  %``Analysis and interpretation of new low-energy pi pi scattering data,''
  Eur.\ Phys.\ J.\  C {\bf 24} (2002) 469
  [arXiv:hep-ph/0112088].

\bibitem{yndurain}
 F.~J.~Yndurain, R.~Garcia-Martin and J.~R.~Pelaez,
  %``Experimental status of the pi pi isoscalar S wave at low energy: f0(600)
  %pole and scattering length,''
  arXiv:hep-ph/0701025.
 
\bibitem{kaminski}
 R.~Kaminski, J.~R.~Pelaez and F.~J.~Yndurain,
  %``The pion-pion scattering amplitude. III: Improving the analysis with
  %forward dispersion relations and Roy equations,''
  arXiv:0710.1150 [hep-ph].


\bibitem{MMS}
  U.-G.~Mei{\ss}ner, G.~M\"uller and S.~Steininger,
  %``Virtual photons in SU(2) chiral perturbation theory and electromagnetic
  %corrections to pi pi scattering,''
  Phys.\ Lett.\ B {\bf 406} (1997) 154
  [Erratum-ibid.\ B {\bf 407} (1997) 454]
  [arXiv:hep-ph/9704377].
  %%CITATION = HEP-PH 9704377;%%

\bibitem{gps}
  E.~Gamiz, J.~Prades and I.~Scimemi,
  %``K --> 3pi final state interactions at NLO in CHPT and Cabibbo's  proposal
  %to measure a(0)-a(2),''
  Eur.\ Phys.\ J.\  C {\bf 50} (2007) 405
  [arXiv:hep-ph/0602023].
  %%CITATION = EPHJA,C50,405;%%

\bibitem{radbfgkr}
  M.~Bissegger et al., work in progress.

\bibitem{knechturech}
  M.~Knecht and R.~Urech,
  %``Virtual photons in low energy pi pi scattering,''
  Nucl.\ Phys.\ B {\bf 519} (1998) 329
  [arXiv:hep-ph/9709348].
  %%CITATION = HEP-PH 9709348.%%



\bibitem{nrqft6}
A.~Gall, J.~Gasser, V.~E.~Lyubovitskij and A.~Rusetsky,
Phys.\ Lett.\ B {\bf 462} (1999) 335 [arXiv:hep-ph/9905309].
%%CITATION = HEP-PH 9905309.%%
\bibitem{nrqft7}
J.~Gasser, V.~E.~Lyubovitskij, A.~Rusetsky and A.~Gall,
Phys.\ Rev.\ D {\bf 64} (2001) 016008 [arXiv:hep-ph/0103157].
%%CITATION = HEP-PH 0103157.%%
\bibitem{nrqft8}
J.~Gasser, V.~E.~Lyubovitskij and A.~Rusetsky,
Phys.\ Lett.\ B {\bf 471} (1999) 244 [arXiv:hep-ph/9910438].
%%CITATION = HEP-PH 9910438
\bibitem{nrqft9}
J.~Schweizer,
Phys.\ Lett.\ B {\bf 587} (2004) 33 [arXiv:hep-ph/0401048].
%%CITATION = HEP-PH 0401048.%%  
\bibitem{nrqft10}
J.~Schweizer,
Eur.\ Phys.\ J.\ C {\bf 36} (2004) 483 [arXiv:hep-ph/0405034].
%%CITATION = HEP-PH 0405034.%%
\bibitem{nrqft11}
V.~E.~Lyubovitskij and A.~Rusetsky,
Phys.\ Lett.\ B {\bf 494} (2000) 9 [arXiv:hep-ph/0009206].
%%CITATION = HEP-PH 0009206.%%
\bibitem{nrqft12}
J.~Gasser, M.~A.~Ivanov, E.~Lipartia, M.~Moj\v{z}i\v{s} and A.~Rusetsky, 
Eur.\ Phys.\ J.\ C {\bf 26} (2002) 13 [arXiv:hep-ph/0206068].
%%CITATION = HEP-PH 0206068.%%
\bibitem{nrqft13}
  U.-G.~Mei{\ss}ner, U.~Raha and A.~Rusetsky,
  Eur.\ Phys.\ J.\ C {\bf 35} (2004) 349
  [arXiv:hep-ph/0402261].
  %%CITATION = HEP-PH 0402261.%%
\bibitem{nrqft14}
  U.-G.~Mei{\ss}ner, U.~Raha and A.~Rusetsky,
  Eur.\ Phys.\ J.\ C {\bf 41} (2005) 213
  [Erratum-ibid.\ C {\bf 45} (2006) 545]
  [arXiv:nucl-th/0501073].
  %%CITATION = NUCL-TH 0501073.%%


\bibitem{tarasovk3pi}
 S.~R.~Gevorkyan, A.~V.~Tarasov and O.~O.~Voskresenskaya,
  %``Electromagnetic corrections to final state interactions in K $\to$ 3pi
  %decays,''
  Phys.\ Lett.\  B {\bf 649} (2007) 159;
  %%CITATION = PHLTA,B649,159;%%
 S.~R.~Gevorkyan, D.~T.~Madigozhin, A.~V.~Tarasov and O.~O.~Voskresenskaya,
  %``Electromagnetic effects and scattering lengths extraction
% from experimental
  %data on K --> 3pi decays,''
  arXiv:hep-ph/0702154.
  %%CITATION = HEP-PH/0702154;%%

\end{thebibliography}
